\documentclass[conference]{IEEEtran}
\IEEEoverridecommandlockouts
\usepackage{cite}
\usepackage{amsmath,amssymb,amsfonts}
\usepackage{algorithmic}
\usepackage{graphicx}
\usepackage{textcomp}
\usepackage{xcolor}

\def\BibTeX{{\rm B\kern-.05em{\sc i\kern-.025em b}\kern-.08em
    T\kern-.1667em\lower.7ex\hbox{E}\kern-.125emX}}
\begin{document}

\title{Radio Environment Maps for Dynamic Frequency Selection in V2X Communications \\
\thanks{The work has been realized within the project no. 2018/29/B/ST7/01241 funded by the National Science Centre in Poland.}
}

\author{\IEEEauthorblockN{Pawe\l~Sroka}
\IEEEauthorblockA{\textit{Poznan University of Technology}\\
Poznan, Poland \\
pawel.sroka@put.poznan.pl}
\and
\IEEEauthorblockN{Pawe\l~Kryszkiewicz}
\IEEEauthorblockA{\textit{Poznan University of Technology}\\
Poznan, Poland \\
pawel.kryszkiewicz@put.poznan.pl}
\and
\IEEEauthorblockN{Adrian~Kliks}
\IEEEauthorblockA{\textit{Poznan University of Technology}\\
Poznan, Poland \\
adrian.kliks@put.poznan.pl}
\and
}

% \IEEEauthorblockN{Micha\l~Sybis}
% \IEEEauthorblockA{\textit{Faculty of Electronics and Telecommunications} \\
% \textit{Poznan University of Technology}\\
% Poznan, Poland \\
% michal.sybis@put.poznan.pl}
\maketitle

\begin{abstract}
In this paper, we investigate the concept of database supported Vehicular Dynamic Spectrum Access (VDSA) for platooning. As various researchers show that the 5.9 GHz band, devoted for Intelligent Transportation Systems, may suffer from congestion of the channel, we propose to offload part of this traffic to white-spaces with the guidance of the active database system. In our work, we describe our measurement campaign which delivered data for population of the dedicated radio environment map. Once the map is created, it was used in three proposed algorithms for VDSA: an optimal and two pragmatic approaches.
\footnote{Copyright © 2020 IEEE. Personal use us permitted. For any other purposes, permission must be obtained from the IEEE by emailing pubs-permissions@ieee.org. This is the author’s version of an article that has been published in the proceedings of 2020 IEEE 91st Vehicular Technology Conference (VTC2020-Spring) and published by IEEE. Changes were made to this version by the publisher prior to publication, the final version of record is available at: http://dx.doi.org/10.1109/VTC2020-Spring48590.2020.9128655. To cite the paper use: P. Sroka, P. Kryszkiewicz and A. Kliks, "Radio Environment Maps for Dynamic Frequency Selection in V2X Communications," 2020 IEEE 91st Vehicular Technology Conference (VTC2020-Spring), 2020, pp. 1-6, doi: 10.1109/VTC2020-Spring48590.2020.9128655.  or visithttps://ieeexplore.ieee.org/document/9128655}
\end{abstract}

\begin{IEEEkeywords}
V2X Communications, Radio Environment Maps, System Coexistence, Dynamic Frequency Selection, Vehicular Dynamic Spectrum Access
\end{IEEEkeywords}

\section{Introduction}
Autonomous driving is one of the possible approaches towards practical realization of the concept of vehicle platooning, i.e, a coordinated movement of a group of vehicles. This group of short-distanced vehicles, typically trucks, forms a convoy, which is led by a platoon leader, responsible for sending steering information to the platoon members. In consequence, mutual exchange of control and steering information within the group enables the autonomous movement of the whole platoon with no human intervention. It is worth noting that the increased interest in platooning is driven by the expected potential revenues. In \cite{EChen_2012}, it was observed that the fuel savings may reach up to 7 or even 15 \% for trucks travelling behind the platoon leader. Such savings translate immediately into a substantial reduction of carbon footprint.\\ 
From the perspective of data exchange between the platoon members and between the platoon and other devices, it is realized by means of short-range wireless communications schemes, such as Dedicated Short-Range Communications (DSRC) or cellular networks (Cellular-V2X, C-V2X).
However, solutions based on IEEE 802.11p and Wireless Access in Vehicular Environment (WAVE) standards, that describe the realization of physical and medium-access layers, may suffer from the prospective medium congestion when the number of communicating cars increases \cite{REDDY18, Bohm13}. In this context, the approach to offload some data to other bands are gaining interest. In our work, we concentrate on utilization of unoccupied television channels (known as TV White Spaces, TVWS \cite{Harrison_2010,Beek_2012}), following the concept of Vehicular Dynamic Spectrum Access (VDSA) \cite{Chen_2011,Chen_2012}. One may observe that the location of the Digital Terrestrial Television (DTT) transmitters as well as the assignment of TV channels to the certain transmission points is relatively stable in long time-scale. Hence, it may be feasible to construct a dedicated database subsystem, which will support VDSA algorithm in selection of appropriate TV band for data transmission. A part of this system is the so-called Radio Environment Map (REM) \cite{Wei_2013}, a dedicated database populated with information describing the spectrum awareness in certain locations. Creation and maintenance of such database is crucial for ensuring high reliability of VDSA decisions. 
In our paper, we concentrate on partial intra-platoon traffic offloading from congested 5.9 GHz band Control Channel (CCH) to TVWSs with support of dedicated REM-based system. The platoon leader is able to select the optimal transmit frequency (within the TV band) based on the proposed REM-supported VDSA algorithm. In order to work on realistic scenario, we have conducted the dedicated measurement campaign (drive test), and populated the REM with measured values of the received DTT signal strength, observed on the TV channels that have to be protected from harmful interference. Investigated VDSA algorithms utilize the knowledge on the observed signal power in TV channels occupied by DTT transmissions, and select such a carrier frequency for data transmission, that minimizes the distortion of DTT transmission and the other platoon transmissions. We have observed that by offloading part of the inter-vehicle traffic to TVWS we can improve the successful reception ratio of intra-platoon messages, thus increasing the reliability of platooning.
The remainder of the paper is organized as follows. First, in section 2, the assumed system model is presented. Next, the measurement campaign is discussed and the process of database population is elaborated. Proposals of the optimized and pragmatic REM-supported VDSA algorithms are given in section 4, whereas the achieved simulation results - in section 5. Finally, the paper is concluded. 

\section{System Model}
\begin{figure}[htbp]
\centerline{\includegraphics[width=0.5\textwidth]{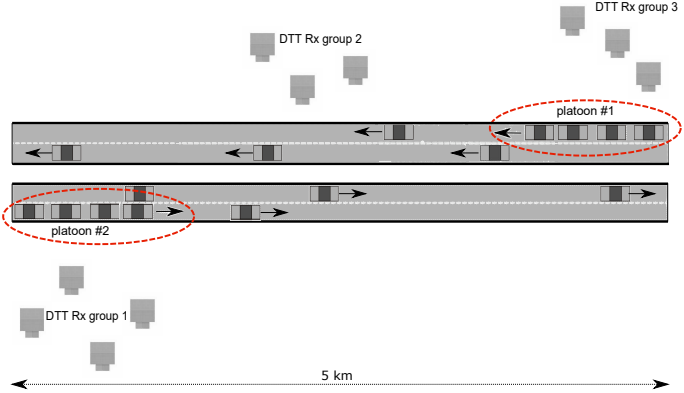}}
\caption{Considered motorway scenario with two platoons and DTT receivers.}
\label{fig_scenario}
\end{figure}

In this work we consider a motorway scenario, such as the one presented in Fig. \ref{fig_scenario}, where multiple platoons of cars travel among other vehicles. We assume that platoon cars are autonomous with their mobility controlled using Cooperative Adaptive Cruise Control (CACC) algorithm proposed in \cite{Raj2000}.\\

In this specific investigation we limit the number of platoons to two, each moving on the outer lane of the motorway (comprising 4 lanes). Each platoon is preceded by a jammer car that periodically reduces its velocity from 130 km/h to 100 km/h and then accelerates back to 130 km/h (single cycle duration is 30 s). Moreover, non-platoon cars are randomly deployed on the inner motorway lanes, with their density equal to 20 cars/km/lane.\\
Platooning using CACC is supported with wireless communications between platoon cars, where each vehicle obtains the needed information from the preceding car and from the platoon leader. We assume that all cars use DSRC according to the IEEE 802.11p standard \cite{IEEE80211} to broadcast Cooperative Awareness Messages (CAMs) comprising basic information on mobility parameters. Non platoon cars transmit CAMs every 100 ms in the CCH at 5.9 GHz. Platoon cars, apart from transmitting CAMs in CCH, have also the possibility to send dedicated CACC messages in additional frequency bands, such as TVWS. When TVWS are used, the CAMs  frequency is reduced to 5Hz, but CACC transmission frequency is also set to 5Hz (so in total the messaging frequency equals 10Hz).\\
The use of TVWS for platooning purposed is limited by the presence of DTT receivers near the motorway that need to be protected (DTT is considered a primary system). These are grouped in three areas, with the exact longitudinal positions and distance to motorway given in Table \ref{tab_DTTpos}.\\
\begin{table}[htbp]
\caption{DTT receivers' positions}
\label{tab_DTTpos}
\begin{tabular}{|c|c|c|c|}
\hline
 DTT Rx id & long. position [m] & dist. to motorway [m] & Group \\
 \hline
 1 & 240 & 120 & 1 \\
 \hline
 2 & 4520 & 164 & 3 \\
 \hline
 3 & 4320 & 244 & 3 \\
 \hline
 4 & 320 & 45 & 1 \\
 \hline
 5 & 1687 & 80 & 2 \\
 \hline
 6 & 4112 & 304 & 3 \\
 \hline
 7 & 632 & 140 & 1 \\
 \hline
 8 & 485 & 270 & 1 \\
 \hline 
 9 & 1463 & 154 & 2 \\
 \hline
 10 & 2087 & 127 & 2 \\
 \hline
\end{tabular}
\end{table}
VDSA algorithm is used to support intra-platoon communications (working as a secondary system) in order to keep the interference to the primary system in TVWS at an acceptable level and to maximize the Signal-to-Interference-and-Noise Ratio (SINR) of the V2V transmission. Hence, a centralized management approach is considered, where the decisions on dynamic spectrum access are supported with REM databases.
\section{Spectrum Measurements and REM Creation}
\begin{figure}[htbp]
\centerline{\includegraphics[width=0.5\textwidth]{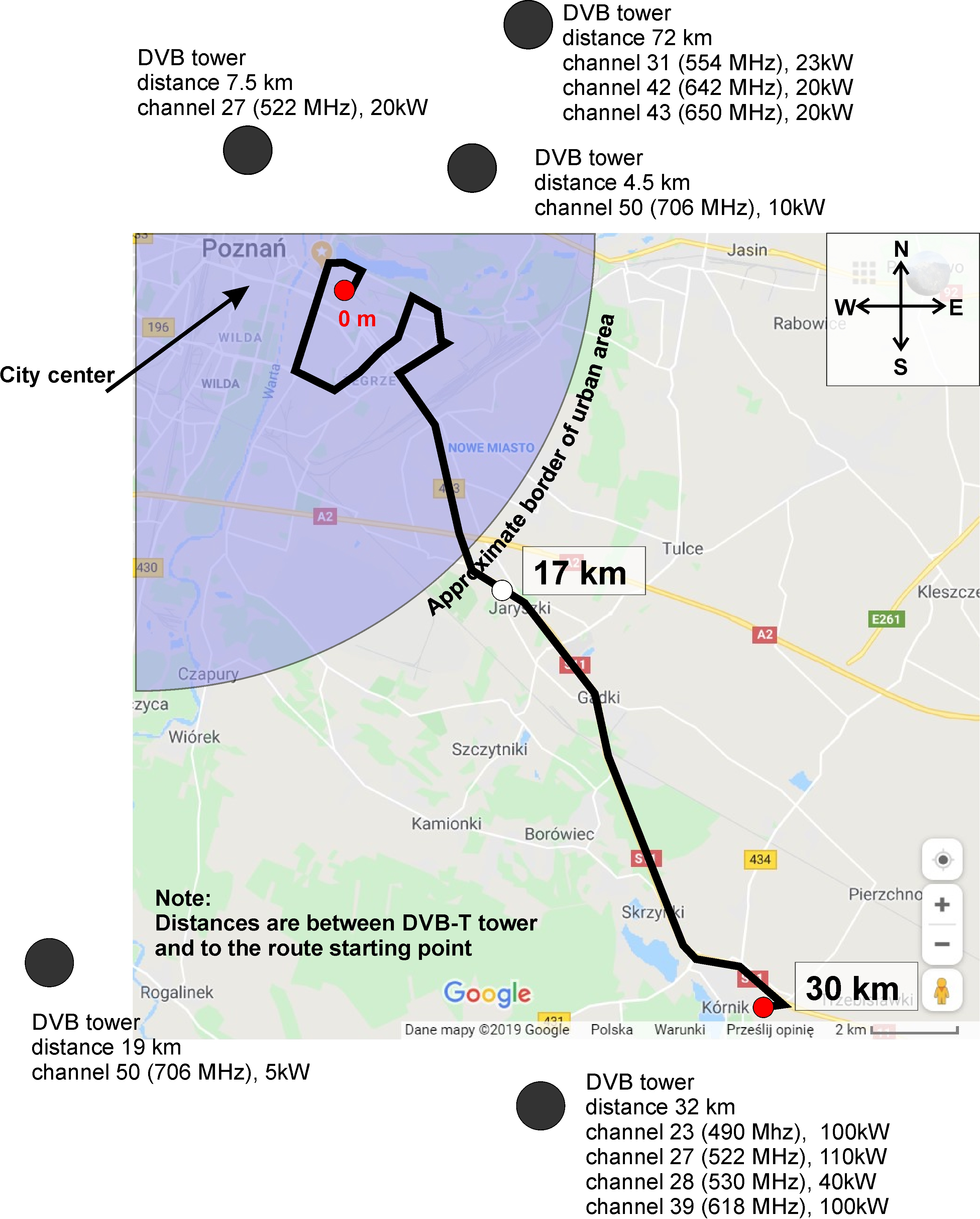}}
\caption{Route for drive test measurements around Poznan (52.4145, 16.9230)}
\label{fig_route}
\end{figure}

\begin{figure}[htbp]
\centerline{\includegraphics[width=0.5\textwidth]{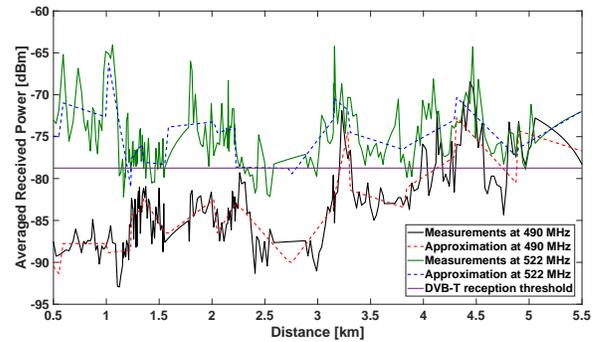}}
\caption{Received DVB-T signal power along the route at channel 23 and 27; solid horizontal line represents the minimum required DVB-T reception threshold}
\label{fig_rec_power}
\end{figure}

The ultimate goal of our analysis is to efficiently utilize the white spaces for data transmission within the platoons with guaranteeing no harm to any potentially existing transmission in the primary network. In our case, we have selected the TV band (470 MHz - 692 MHz) for detailed evaluation, thus the protection of any DTT receiver is of highest importance. In order to work on real-data, the dedicated  measurement campaign has been conducted, when the drive test along the route shown in Fig.~\ref{fig_route} has been made. 
Based on the freely available coverage maps for Digital Video Broadcasting - Terrestrial (DVB-T) transmission around Poznań (in Poland), where the experiment has been carried out, eight TV channels have been specified in which DTT signal should be strong enough to allow for successful reception, i.e., the received power is above -78.75 dB \cite{DSA_ruler_DTT_2017}. These are channels numbered 23, 27, 28, 31, 39, 42, 43 and 50, and the azimuths to these DVB-T towers are shown in Fig.~\ref{fig_route}. From the perspective of the secondary transmission, these channels may be considered as potentially (in certain locations) occupied by the DTT transmission, and as such these should be protected. For our investigation, we have arbitrarily selected the band between 486 MHZ - 526 MHz, i.e., two outer TV channels are occupied by DTT transmission (i.e., channels 23 and 27 with corresponding centre frequency equal to 490 MHz, and 522 MHz), whereas the middle channels may be stated as potentially free (i.e., the centre frequencies are equal to 498 MHz, 506 MHz, and 514 MHz). For these two outer channels, during the drive tests we have measured the received signal strength, first, by collecting in-phase and quadrature (IQ) samples, and second, by computing received power at each consecutive location on the route. These samples have been collected by means of Rohde\&Schwarz FLS6 spectrum analyzer equipped with AOR DA753 aerial, and saved in laptop. The antenna has been mounted to the car railing, and connected through short jumper (H155 cable) with the analyzer deployed in the car interior. The entire setup has been also supplied with power delivered through 12 Volts car current sockets. The precise location of the car has been obtained by means of the dedicated GPS receiver, which delivered the coordinates to the laptop. 
The entire measurement process has been controlled by the software run in Matlab environment. At each location 523776 IQ samples\footnote{This number was selected as the maximum available in the spectrum analyzer for a given sampling frequency.} have been collected with sampling frequency set to $\frac{64}{7}$ Msps, and one resultant averaged value of received power was computed. As the average speed of the car was around 60 km/h, these samples have been collected over around 1s that corresponds to the distance of around 1 m. Such a procedure has been repeated 50 times, and such a bunch of results have been stored into appropriate files. As the saving process takes some seconds, there are some small gaps in measurements, however, the missing values have been linearly interpolated. Achieved measurement results of the received DVB-T power in two channels (23 and 27) are shown in Fig.~\ref{fig_rec_power} as a function of route distance. Moreover, the DVB-T reception threshold (i.e. -78.75 dBm) is also included in the figure - when the received power of the DVB-T signal is below the violet line, it is impossible to decode the DVB-T signal. Let us notice that we have intentionally selected such fragment of our route where there are significant changes of the received signal power, and where we observe crossing of the threshold line. 
In general, in the context of REM creation, such drive tests as described above shall be repeated many times to obtain  statistically correct representation considering possible changes in the environment. One of important aspects is the correlation of the received power in space. While pathloss from DTT transmitter to adjacent sensing points should be similar, some variation caused by shadowing is possible.
Thus, we decided to approximate the changes in the observed received power by means of the polynomial function minimizing mean square error of approximation. In particular, we split the entire route into segments, each containing 20 measurement locations (what corresponds to approximately 600 m in distance for average speed of 60 km/h). For each segment, the first order polynomial was fitted, with the achieved polylines shown in dashed style in Fig.~\ref{fig_rec_power} for both DVB-T channels. Moreover, for each segment the standard deviation of approximation error was calculated and stored in the database as the factor representing shadowing phenomenon. In our experiment, the values of shadowing vary from 1.5 dB to 4.5 dB depending on the location (segment). Finally, in REM we stored the estimated DTT received signal power and the corresponding standard deviation values resulting from shadowing. 

\section{REM-based VDSA Algorithm}
The spectrum allocation for a platoon should guarantee successful packet transmission and reception by interested transmitter/receiver while protecting the incumbent/primary users (PU) of a given spectrum band. Let us define $\mathcal{I}$ as a set of all transmitting/receiving vehicles in all considered platoons. $\mathcal{I}_{\mathrm{RX}}\subset \mathcal{I}$ is a set of receiving vehicles, that are typically all cars except for the platoon leader. Each car out of this set, i.e., $i_{RX}\in \mathcal{I}_{\mathrm{RX}}$, is to receive transmissions from $i_{TX}\in \mathcal{I}_{\mathrm{TX}}(i_{RX})$ transmitting cars. $\mathcal{I}_{\mathrm{TX}}(i_{RX})$ is a set of transmitters that are "listened" by $i_{RX}$, in the considered scenario the platoon leader and the car in front of $i_{RX}$. For safety reasons it is required to maximize the probability that $i_{TX}\in \mathcal{I}_{\mathrm{TX}}(i_{RX})$ is allowed to transmit and $i_{RX}$ manages to receive this transmission, i.e., $\max Pr(i_{TX} \cap i_{RX})$ for $i_{TX}\in \mathcal{I}_{\mathrm{TX}}(i_{RX})$. It can be assumed that any packet reception is similarly important for safe platoon operation. As such, minimal probability over all pairs of $i_{RX}$ and $i_{TX}\in \mathcal{I}_{\mathrm{TX}}(i_{RX})$ should be maximized:
\begin{equation}
    \max \min_{\substack{i_{RX}\\ i_{TX}\in \mathcal{I}_{\mathrm{TX}}(i_{RX})}} Pr(i_{TX} \cap i_{RX}).
\end{equation}
Using conditional probability definition $Pr(i_{TX} \cap i_{RX})$ can be factorized into
\begin{equation}
 Pr(i_{TX} \cap i_{RX})=Pr(i_{RX}\mid i_{TX})Pr(i_{TX})
\end{equation}
where $Pr(i_{TX})$ is the probability that $i_{TX}$ starts transmission, i.e., Carrier-Sense Multiple Access with Collision Avoidance (CSMA-CA) protocol detects channel as unoccupied over sufficient period of time, and $Pr(i_{RX}\mid i_{TX})$ is the probability of successful reception at node $i_{RX}$ while $i_{TX}$ is transmitting. $Pr(i_{TX})$ is a function of threshold used to assess channel as unoccupied. It can be always optimized in such a way that each transmitter manages to transmit its frame before new frame is ready to be transmitted. In the ultimate configuration the threshold can be set to infinity meaning that each transmitter releases its frame immediately after it enters its buffer, i.e., pure ALOHA system is obtained. However, maximization of reception rate, i.e., maximization of $Pr(i_{RX}\mid i_{TX})$, is obtained if maximum SINR is obtained at $i_{RX}$ for transmission from $i_{TX}$, which is denoted as
\begin{equation}
SINR_{i_{RX},i_{TX}}=\frac{P_{i_{TX}}\left|h_{i_{TX},i_{RX}}\right|^{2}}{N+I^{PU-V}_{i_{RX}}(f_{Q(i_{TX})})+I^{V-V}_{i_{RX},i_{TX}}
}
\end{equation}
where $P_{i_{TX}}$ is the transmit power of $i_{TX}$ transmitter, $\left|h_{i_{TX},i_{RX}}\right|^{2}$ is the wireless channel propagation gain (including antenna gains) between $i_{TX}$ transmitter and $i_{RX}$ receiver, $N$ is the thermal noise power at the receiver,
$I^{PU-V}_{i_{RX}}(f_{Q(i_{TX})})$ is the interference power observed at $i_{RX}$ receiver from the primary users utilizing bands of interest and $I^{V-V}_{i_{RX}}$ is the interference power from all cars in the platoons. The function $Q(~)$ maps the index of vehicle to the platoon index $k$ and $f_k$ is the center frequency used by $k-th$ platoon.  
The interference power $I^{PU-V}_{i_{RX}}(f_{Q(i_{TX})})$ can be a random variable as a result of variation in PU transmission (e.g., bursty transmission) or variation in pathloss. As for the optimization of platoons parameters, an estimate of $I^{PU-V}_{i_{RX}}(f_{Q(i_{TX})})$ at location of $i_{RX}$ receiver at frequency $f_{Q(i_{TX})}$  can be obtained using the parameters of the transmitter and the pathloss, utilizing some propagation model. In the case of REM-driven optimization with spectrum sensing capable devices (e.g., roadside units or vehicles) local PU measurements can become available. In the case of DTT being the PU system, the measured power of $i_{DTT}$ DTT channel at the location of $i_{RX}$, i.e., $P^{DTT}_{i_{DTT},i_{RX}}$, is relatively fixed in time. As such the interference observed by $i_{RX}$ receiver depends on Adjacent Channel Interference Ratio (ACIR) between DTT transmission at frequency $f^{DTT}_{i_{DTT}}$ and platoon reception at frequency $f_{Q(i_{TX})}$ denoted as $ACIR^{DTT-V}(f^{DTT}_{i_{DTT}}-f_{Q(i_{TX})})$. $ACIR^{DTT-V}(f^{DTT}_{i_{DTT}}-f_{Q(i_{TX})})\in \langle 0;1\rangle$ is a linear value describing coupling between transmission and reception considering both emitted signal spectrum (described typically by Adjacent Channel Leakage Ratio or spectrum emission mask) and ability of the receiver to reject adjacent channel signal (described typically by Adjacent Channel Selectivity). As such we can define
\begin{equation}
I^{PU-V}_{i_{RX}}=\sum_{i_{DTT}}
P^{DTT}_{i_{DTT},i_{RX}}
ACIR^{DTT-V}(f^{DTT}_{i_{DTT}}-f_{Q(i_{TX})}).
\end{equation}
The interference from other vehicles is a random variable depending on a random event of $j_{TX}\neq i_{TX}$ transmitting during $i_{TX}$ transmission. The simplest approach is to use the expectation of interference power as 
\begin{align}
I^{V-V}_{i_{RX},i_{TX}}=&\sum_{j_{TX}\neq i_{TX}}
Pr\left(j_{TX}\mid i_{TX}\right)P_{j_{TX}}
\left|h_{j_{TX},i_{RX}}\right|^{2}
\nonumber
\\&ACIR^{V-V}(f_{Q(j_{TX})}-f_{Q(i_{TX})})
\label{eq_int_v_v_basic}
\end{align}
where $ACIR^{V-V}(f_{Q(j_{TX})}-f_{Q(i_{TX})})\in \langle 0; 1\rangle$ is ACIR in linear scale between wanted transmission on frequency $f_{Q(i_{TX})}$ and interfering transmission on frequency $f_{Q(j_{TX})}$ and $Pr\left(j_{TX}\mid i_{TX}\right)$ is the probability that $j_{TX}$ transmits while $i_{TX}$ is transmitting. This conditional probability depends on the parameters of CSMA-CA protocol, e.g., the threshold for assessment of a given channel as free or occupied $\Gamma$. In the case of pure ALOHA protocol or when the transmission threshold is set to infinity, both transmissions are independent meaning $Pr\left(j_{TX}\mid i_{TX}\right)=Pr\left(j_{TX}\right) Pr\left(i_{TX}\right)/Pr\left(i_{TX}\right)=Pr\left(j_{TX}\right)$. The probability $Pr\left(j_{TX}\right)$ can be calculated as a period of time when the wireless channel is occupied by $j_{TX}$ transmission divided by the period of time between the two transmission requests of $j_{TX}$. 
In practical system, because of CSMA-CA operation the transmitters other than $i_{TX}$ within its "interference range" should remain silent. The most probable is interference from the "hidden nodes", i.e., $\mathcal{I}_{TX}^{HN}(i_{TX})=\{j_{TX}: j_{TX}\in \mathcal{I} \land P_{i_{TX}}
\left|h_{i_{TX},j_{TX}}\right|^{2}
ACIR^{V-V}(f_{Q(j_{TX})}-f_{Q(i_{TX})})<\Gamma \}$. Observe, that the definition of set $\mathcal{I}_{TX}^{HN}(i_{TX})$ does not consider mutual interference from many sources for simplicity. Therefore, (\ref{eq_int_v_v_basic}) can be approximated as
\begin{align}
I^{V-V}_{i_{RX}}=\!\!\!\!\!\!\!\sum_{j_{TX}\in \mathcal{I}_{TX}^{HN}(i_{TX})
}\!\!\!\!\!\!\! &
Pr\left(j_{TX}\right)P_{j_{TX}}
\left|h_{j_{TX},i_{RX}}\right|^{2}
\nonumber \\ &
ACIR^{V-V}(f_{Q(j_{TX})}-f_{Q(i_{TX})})
\label{eq_int_v_v_approx}
\end{align}
\subsection{Protection of Primary Users} 
As the platoons are to operate in a spectrum band designated for other wireless systems, some protection of the incumbents is required. The requirements can vary from system to system. In the case of dynamic access to DTT spectrum a lot of research was done and reported, e.g., by regulators \cite{OFCOM_TVWS_2015} or some standardization organizations \cite{DSA_ruler_DTT_2017}. The highest spectral efficiency can be achieved when the location of protected DTT receivers is known. If it is not known, worst case assumption is to be made. According to \cite{DSA_ruler_DTT_2017} DTT RX distanced by 60 m from a transmitting vehicle should be assumed to be able to receive $i_{DTT}$ DTT channel if the SNR is above a certain threshold. Although this metric can be estimated using some propagation model \cite{DSA_ruler_DTT_2017}, in our case sensing data from vehicles, stored in REM can be utilized. It can be assumed that propagation conditions to the DTT receiver located in a close distance to motorway are similar to the one to the $i_{TX}$ transmitter. As such $i_{DTT}$ should be protected if $P^{DTT}_{i_{DTT},i_{TX}}>\Gamma_{DTT}$. As for \cite{DSA_ruler_DTT_2017} $\left(\Gamma_{DTT}\right)_{dB}=-78.75 dBm$. The interference generated by $i_{TX}$ node transmission has to guarantee Signal-to-Interference Ratio (SIR) at DTT RX is above certain threshold, i.e.,
\begin{align}
   & \forall{i_{DTT},i_{TX}:P^{DTT}_{i_{DTT},i_{TX}}>\Gamma_{DTT}}
    \nonumber \\ &
    P_{i_{TX}}|h_{i_{TX}}^{DTT}|^2 ACIR^{V-DTT}(f^{DTT}_{i_{DTT}}-f_{Q(i_{TX})},P^{DTT}_{i_{DTT},i_{TX}})
    \nonumber \\ &
    <P^{DTT}_{i_{DTT},i_{TX}}/SIR_{min}^{DTT},
\end{align}
where $|h_{i_{TX}}^{DTT}|^2$ is channel gain between $i_{TX}$ transmitter and a potential DTT RX (in this work we assume that the DTT power at the protected receivers listed in Table \ref{tab_DTTpos} is available in REM), $ACIR^{V-DTT}(~,~)$ is a linear ACIR between platoon transmission and DTT reception dependent on the difference in carrier frequency and the received power of DTT signal (according to \cite{OFCOM_TVWS_2015} the DTT receiver saturates for stronger DTT signal reducing selectivity) and $SIR_{min}^{DTT}$ is the minimum required SIR for DTT signal and platoon-based interference. According to \cite{DSA_ruler_DTT_2017} the value of 39.5 dB can be used. 
\subsection{ Problem formulation and solution}
Based on the previous derivations the dynamic frequency selection optimization problem can be defined as
\begin{align}
  &  \max_{f_k} \min_{\substack{i_{RX}\\ i_{TX}\in \mathcal{I}_{\mathrm{TX}}(i_{RX})}}
    SINR_{i_{RX},i_{TX}}
\nonumber \\ &
s.t. \forall{i_{DTT},i_{TX}:P^{DTT}_{i_{DTT},i_{TX}}>\Gamma_{DTT}}
    \nonumber \\ &
    P_{i_{TX}}|h_{i_{TX}}^{DTT}|^2 ACIR^{V-DTT}(f^{DTT}_{i_{DTT}}-f_{Q(i_{TX})},P^{DTT}_{i_{DTT},i_{TX}})
    \nonumber \\ &
    <P^{DTT}_{i_{DTT},i_{TX}}/SIR_{min}^{DTT}.
\end{align}
Assuming there is a limited set of possible frequency channels to be utilized by platoons, optimization of $f_k$ can be carried by an exhaustive search algorithm. Alternatively, more pragmatic approach can be considered, where the available frequency range for dynamic selection is calculated first (based on the DTT protection constraint), with the inter-platoon interference minimized in the second step by choosing the centre frequencies set providing maximum distance between the selected frequencies of all platoons (DTT to V2V interference is not accounted for in selection of centre frequencies).

\section{Simulation Results}
\begin{figure}[htbp]
\centerline{\includegraphics[width=0.5\textwidth]{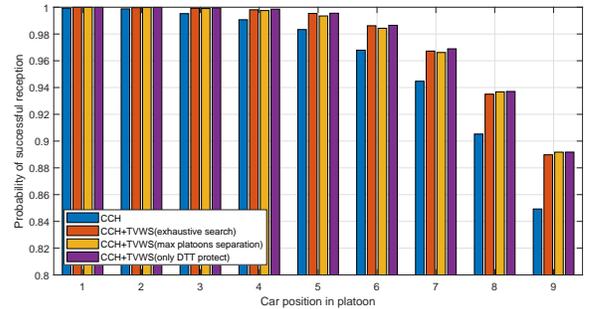}}
\caption{Probability of successful reception of leaders' packets.}
\label{fig_leaderP}
\end{figure}

To evaluate the gains from dynamic spectrum access for platooning using the TVWS we performed system-level simulations of the considered scenario using a simulation tool developed in C++, that was calibrated and used also in work described in \cite{VUK2018,Syb19}. The duration of a single simulation run was set to 140 s, which is equivalent to the platoon traveling almost 5 km. The assumed DTT protection levels were the following: the minimum DTT receive power to be protected was set to -80 dBm (to include a small gap from the value proposed in section III.A with the aim to compensate for shadowing); for the protected DTT receivers the minimum required SIR level was set to 39.5 dB. Three strategies of dynamic selection of centre frequencies were considered:
\begin{itemize}
    \item Exhaustive search according to the rationale presented in section IV.B - (8).
    \item Protection of DTT receivers according to (7) and then maximizing the distance in frequency between different platoons (denoted later as \em{Max platoon separation}\em{}).
    \item Only protection of DTT receivers according to (7) (denoted as \em{DTT protect}\em{}).
\end{itemize}
Presence of two DTT bands was accounted for (with centre frequencies at 490 MHz and 522 MHz), hence, the considered centre frequencies range for platooning was set between 499 MHz and 513 MHz to avoid full overlap with the DTT. The frequency selection was performed periodically every 1 s.\\
The most important aspect of intra-platoon communications is the reliability of the transmission - the assumed target reception rate of messages should be over 99\% to reliably perform platooning. Such requirement is problematic when only CCH is used, as stated in \cite{VUK2018, Syb19} and shown in Fig.~\ref{fig_leaderP}, where the reception rate of leader packets is given for the consecutive car positions in platoon. We do not present the results of transmission of the preceding car messages as reception rate in this case is close to 1 due to small distances between the transmitter and the receiver.\\
The use of TVWS for transmission of CACC packets improves the reception rate (see Fig. \ref{fig_leaderP}) due to two factors: reduced messaging rate in CCH (reduced congestion) and increased transmission range (use of lower frequencies). The correct reception probability is almost the same for all three methods using TVWS. However, one should note that the reception rate with dynamic frequency selection is still below the required level, which is mostly due to the interference from DTT.\\

\begin{figure}[htbp]
\centerline{\includegraphics[width=0.5\textwidth]{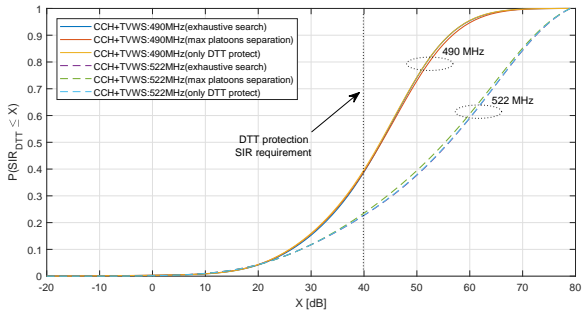}}
\caption{Cumulative distribution of SIR for the protected DTT receivers.}
\label{fig_dttSIR}
\end{figure}

The intra-platoon communications in TVWS as secondary system should not cause the degradation of DTT service. Hence, an important aspect is to keep the SIR level of DVB-T transmission above the required threshold of 40 dB. Fig. \ref{fig_dttSIR} shows the SIR cumulative distribution obtained in simulations for the receivers that required protection (received signal power above -80 dBm). One can note that for all three considered strategies the results are similar, with the exhaustive search performing slightly better. However, for both considered DTT bands a significant fraction of SIR samples fall below the requirement threshold. For the 490 MHz band in almost 40\% of cases the V2V transmission causes unacceptable service degradation. Similar situation can be observed for the 522 MHz band, where in about 22\% of cases the SIR level is below the required one. This indicates that the available spectrum is probably too narrow to effectively mitigate the interference to DTT. Moreover, the DTT power levels observed in the considered scenario are very challenging, as the lower the DTT power that needs to be protected (the closer it is to the -80 dB threshold) the lower the V2V interference power should be. This is also reflected by the difference in results between the 490 MHz and 522 MHz bands. For 522 MHz band the received DTT power is typically higher than for the 490 MHz band (see Fig. \ref{fig_rec_power}), thus, it is easier to protect it. Hence, from the DTT protection point of view, the ideal situation would be if the DTT received power is very high.\\

Apart from the interference introduced to the primary system, another cost of VDSA is the need to disseminate the information and coordinate the frequency changes in the whole platoon. The impact of this factor depends on the average number of frequency switches performed by each platoon, presented in Table \ref{tab_freqSwitch}. One can note that the number of frequency changes is much higher in case of platoon~1. This is due to the fact that there are several DTT receivers to be protected that are much closer to platoon 1~than to platoon~2. It can be also observed, that the lowest number of switches is noted with the exhaustive search algorithm.\\

\begin{table}[htbp]
\caption{Average number of frequency changes}
\label{tab_freqSwitch}
\begin{tabular}{|c|c|c|c|}
\hline
Platoon & Exhaustive search & Max platoons separation & DTT protect \\
\hline
1 & 35.4 & 36.34 & 40.42 \\
\hline
2 & 7.38 & 6.42 & 8.04 \\
\hline
\end{tabular}
\end{table}
Concluding, the dynamic use of TVWS has significant potential to improve the intra-platoon communications and, consequently, the reliability of CACC. However, in order to achieve the gains and fulfil the constraints on the  primary system protection, a large set of possible frequency ranges is required. Simulation results showed that 24 MHz span is not enough to effectively protect DTT, and the performance of all algorithms is severely limited by the protection constraints (all strategies, despite the obvious differences, performed similarly). Finally, the dynamic centre frequency selection can be complemented with the transmit power adaptation to provide additional degree of freedom in the optimization problem.

%\begin{figure}[htbp]
%\centerline{\includegraphics[width=0.5\textwidth]{figures/latency_mod.png}}
%\caption{Cumulative distribution of CACC packets' latency.}
%\label{fig_latency}
%\end{figure}

\section{Conclusions}
In this paper we have presented the idea of REM-supported VDSA using TVWS bands for intra-platoon V2V communications. The dynamic centre frequency selection problem has been formulated, including the constraints aiming at protection of the primary system and the optimization of the performance of V2V transmission. Results of computer simulations, aided with measurements of real DTT power pattern in space and frequency, showed the potential, but also the challenges of the dynamic REM-supported use of TVWS for intra-platoon communications.

\
\bibliographystyle{IEEEtran}
\bibliography{bibtex.bib}

\begin{thebibliography}{10}
\providecommand{\url}[1]{#1}
\csname url@samestyle\endcsname
\providecommand{\newblock}{\relax}
\providecommand{\bibinfo}[2]{#2}
\providecommand{\BIBentrySTDinterwordspacing}{\spaceskip=0pt\relax}
\providecommand{\BIBentryALTinterwordstretchfactor}{4}
\providecommand{\BIBentryALTinterwordspacing}{\spaceskip=\fontdimen2\font plus
\BIBentryALTinterwordstretchfactor\fontdimen3\font minus
  \fontdimen4\font\relax}
\providecommand{\BIBforeignlanguage}[2]{{%
\expandafter\ifx\csname l@#1\endcsname\relax
\typeout{** WARNING: IEEEtran.bst: No hyphenation pattern has been}%
\typeout{** loaded for the language `#1'. Using the pattern for}%
\typeout{** the default language instead.}%
\else
\language=\csname l@#1\endcsname
\fi
#2}}
\providecommand{\BIBdecl}{\relax}
\BIBdecl

\bibitem{EChen_2012}
E.~{Chen}, ``{Overview of the SARTRE Platooning Project: Technology Leadership
  Brief},'' \emph{SAE International}, p.~4, October 2012.

\bibitem{REDDY18}
\BIBentryALTinterwordspacing
R.~G. Rajeswar and R.~Ramanathan, ``{An Empirical study on MAC layer in IEEE
  802.11p/WAVE based Vehicular Ad hoc Networks},'' \emph{Procedia Computer
  Science}, vol. 143, pp. 720 -- 727, 2018. [Online]. Available:
  \url{http://www.sciencedirect.com/science/article/pii/S1877050918321410}
\BIBentrySTDinterwordspacing

\bibitem{Bohm13}
A.~{Böhm}, M.~{Jonsson}, and E.~{Uhlemann}, ``Performance comparison of a
  platooning application using the ieee 802.11p mac on the control channel and
  a centralized mac on a service channel,'' in \emph{2013 IEEE 9th
  International Conference on Wireless and Mobile Computing, Networking and
  Communications (WiMob)}, Oct 2013, pp. 545--552.

\bibitem{Harrison_2010}
K.~{Harrison}, S.~M. {Mishra}, and A.~{Sahai}, ``How much white-space capacity
  is there?'' in \emph{2010 IEEE Symposium on New Frontiers in Dynamic Spectrum
  (DySPAN)}, April 2010, pp. 1--10.

\bibitem{Beek_2012}
J.~{van de Beek}, J.~{Riihijarvi}, A.~{Achtzehn}, and P.~{Mahonen}, ``Tv white
  space in europe,'' \emph{IEEE Transactions on Mobile Computing}, vol.~11,
  no.~2, pp. 178--188, Feb 2012.

\bibitem{Chen_2011}
S.~{Chen}, A.~M. {Wyglinski}, S.~{Pagadarai}, R.~{Vuyyuru}, and O.~{Altintas},
  ``Feasibility analysis of vehicular dynamic spectrum access via queueing
  theory model,'' \emph{IEEE Communications Magazine}, vol.~49, no.~11, pp.
  156--163, November 2011.

\bibitem{Chen_2012}
S.~{Chen}, R.~{Vuyyuru}, O.~{Altintas}, and A.~M. {Wyglinski}, ``Learning-based
  channel selection of vdsa networks in shared tv whitespace,'' in \emph{2012
  IEEE Vehicular Technology Conference (VTC Fall)}, Sep. 2012, pp. 1--5.

\bibitem{Wei_2013}
Z.~{Wei}, Q.~{Zhang}, Z.~{Feng}, W.~{Li}, and T.~A. {Gulliver}, ``On the
  construction of radio environment maps for cognitive radio networks,'' in
  \emph{2013 IEEE Wireless Communications and Networking Conference (WCNC)},
  April 2013, pp. 4504--4509.

\bibitem{Raj2000}
R.~{Rajamani}, {Han-Shue Tan}, {Boon Kait Law}, and {Wei-Bin Zhang},
  ``Demonstration of integrated longitudinal and lateral control for the
  operation of automated vehicles in platoons,'' \emph{IEEE Transactions on
  Control Systems Technology}, vol.~8, no.~4, pp. 695--708, July 2000.

\bibitem{IEEE80211}
``{IEEE} {S}tandard for {I}nformation technology--{T}elecommunications and
  information exchange between systems local and metropolitan area
  networks--{S}pecific requirements -- {P}art 11: {W}ireless {LAN} {M}edium
  {A}ccess {C}ontrol ({MAC}) and {P}hysical {L}ayer ({PHY}) {S}pecifications,''
  Dec. 2016.

\bibitem{DSA_ruler_DTT_2017}
\BIBentryALTinterwordspacing
{Dynamic Spectrum Alliance}, ``Model rules and regulations for the use of
  television white spaces v2.0,'' 2017. [Online]. Available:
  \url{http://dynamicspectrumalliance.org/wp-content/uploads/2018/01/Model-Rules-and-Regulations-for-the-use-of-TVWS.pdf}
\BIBentrySTDinterwordspacing

\bibitem{OFCOM_TVWS_2015}
\BIBentryALTinterwordspacing
{OFCOM}, ``Implementing tv white spaces,'' 2015. [Online]. Available:
  \url{{https://www.ofcom.org.uk/\_\_data/assets/pdf\_file/0025/58921/annexes.pdf}}
\BIBentrySTDinterwordspacing

\bibitem{VUK2018}
\BIBentryALTinterwordspacing
V.~Vukadinovic \emph{et~al.}, ``{3GPP C-V2X} and {IEEE 802.11p} for
  vehicle-to-vehicle communications in highway platooning scenarios,'' \emph{Ad
  Hoc Networks}, vol.~74, pp. 17 -- 29, 2018. [Online]. Available:
  \url{http://www.sciencedirect.com/science/article/pii/S157087051830057X}
\BIBentrySTDinterwordspacing

\bibitem{Syb19}
M.~{Sybis} \emph{et~al.}, ``Communication aspects of a modified cooperative
  adaptive cruise control algorithm,'' \emph{IEEE Transactions on Intelligent
  Transportation Systems}, pp. 1--11, 2019.

\end{thebibliography}

\end{document}